\documentclass[aps, prd, preprint, preprintnumbers, showpacs,
showkeys, amsfonts, nofootinbib,floatfix,12]{revtex4-1}
 \usepackage{graphicx}
\usepackage{epsfig}
\usepackage{rotate}
\usepackage{rotating}
\usepackage{multirow}
\usepackage[T1]{fontenc}
\usepackage{array}
\usepackage{graphicx,
}
\usepackage{epstopdf}
\usepackage{amsmath,amssymb,hyperref,enumerate}
\usepackage{pstricks}
\usepackage{slashed}
\usepackage{url}
\usepackage{color}
\usepackage{textcomp}
\hypersetup{colorlinks,citecolor= nicegreen,linkcolor= nicered}
\definecolor{nicered}{rgb}{0.7,0.1,0.1}
\definecolor{nicegreen}{rgb}{0.1,0.5,0.1}

\usepackage[normalem]{ulem}

\def\({\left(}
\def\){\right)}
\def\[{\left[}
\def\]{\right]}

\newcommand{\simgt}{\lower.7ex\hbox{$\;\stackrel{\textstyle>}{\sim}\;$}}
\newcommand{\simlt}{\lower.7ex\hbox{$\;\stackrel{\textstyle<}{\sim}\;$}}

\begin{document}
\vspace{-0.5cm}
\preprint {RECAPP-HRI-2014-023; CUMQ/HEP 186, HIP-2014-20/TH}
\vspace{1.2cm}


\title {Probing the light radion through diphotons at \\the Large Hadron Collider}


\author {Satyaki Bhattacharya$^{(1)}$} \author{Mariana
  Frank$^{(2)}$}
\author{Katri Huitu$^{(3)}$}
\author {Ushoshi Maitra$^{(4)}$}
\author {Biswarup Mukhopadhyaya$^{(4)}$} \author{Santosh Kumar
  Rai$^{(4)}$} \affiliation{$^{(1)}$
  Saha Institute of Nuclear Physics, Sector-I, Block-AF, Bidhannagar,
  Kolkata 700064, India,} \affiliation{$^{(2)}$Department of Physics,
  Concordia University, 7141 Sherbrooke St. West, Montreal, Quebec,
  Canada H4B 1R6,} \affiliation{$^{(3)}$Department of Physics and
  Helsinki Institute of Physics, P.O. Box 64 (Gustaf H\"allstr\"omin
  katu 2), FIN-00014 University of Helsinki, Helsinki, Finland,}
\affiliation{$^{(4)}$Regional Centre for Accelerator-based Particle
  Physics, Harish-Chandra Research Institute, Chhatnag Road, Jhusi,
  Allahabad 211019, India.}


\begin{abstract}
A radion in a scenario with a warped extra dimension can be lighter
than the Higgs boson, even if the Kaluza-Klein excitation 
modes of the graviton turn out to be in the multi-TeV region. The discovery of such a light radion would be gateway to new physics.
We show how the two-photon mode of decay can enable us to probe
a radion in the mass range 60 - 110~GeV.  We take into account
the diphoton background, including fragmentation effects, and include cuts designed to
suppress the background to the maximum possible extent.   Our conclusion is that,  
with an integrated luminosity of 3000~$fb^{-1}$ or less, the next run of the Large
Hadron Collider should be able to detect a radion in this mass range, 
with a significance of 5 standard deviations or more. 
 \end{abstract}

\maketitle

\section{Introduction}
The Large Hadron Collider (LHC) data from the $\sqrt{s}=7$ and $8$~TeV
runs analyzed by the ATLAS \cite{ATLAS} and CMS \cite{CMS}
Collaborations have more or less confirmed a scalar particle whose properties
agree with those of the Standard Model (SM) Higgs boson. Although more
analyses are needed to confirm it to be purely the SM Higgs with 
exact SM like couplings, the question as to whether the SM is the final
theory is still very much open.

Issues ranging from the naturalness of the Higgs mass
to the dark matter content of the universe suggest physics
beyond the standard model (BSM). This has prompted physicists to
look for new particles or symmetries around the TeV scale.  
The lack of new physics signals at the LHC may carry
the message that we have to seek somewhat
higher scales to see BSM physics, especially if it exists in one of
the currently popular forms. That, in turn, has led to conjectures
about new physics  above 1 TeV, which can still address the
naturalness issue, albeit with some degree of fine-tuning. In the
process, however, a question that is perhaps not being asked with
sufficient emphasis is: could new physics, for a change, lie hidden at
a relatively low mass scale, not yet discovered just because of experimental difficulties
in unraveling it? We address one such instance in this paper.

In this context, one BSM scenario which catches one's imagination  is
one with a warped extra space-like dimension, first proposed by Randall
and Sundrum (RS).  The RS model provides an elegant explanation of the
large hierarchy between the electroweak scale (100~GeV -- 1~TeV) and the
Planck scale ($10^{19}$~GeV) in terms of an exponential damping of the
gravitational field across a small compact fifth dimension, without
invoking unnaturally large numbers \cite{Randall:1999ee}. This is
achieved through a non-factorizable geometry with an exponential warp
factor, whereas the additional spatial dimension is compactified on a
$S_1/Z_2$ topology which corresponds to a once-folded circle, with two
$D_3$-branes sitting at the orbifold fixed points.  The original RS
model is based on the assumption that the SM fields are localized on
one of the $D_3$-branes (called the visible brane, at $y = r_c \pi$,
where $r_c$ is the radius of the compact dimension and $y$ is the
co-ordinate along that dimension) and only gravity propagates in the
bulk. Compactification results in a massless state and a tower of
massive modes of the spin-2 graviton on the visible brane. While
resonant production and decays of the massive graviton are rather
spectacular phenomena, the absence of such signals has pushed the lower
limit on the massive modes to about 2.7~TeV \cite{Aad:2014cka}.

The radius of the extra dimension in the RS model is assumed to be
fixed by a given constant and needs stabilizing against quantum
fluctuations parametrized by a scalar field ($\varphi(x)$), {\it viz.}
the radion.  Goldberger and Wise \cite{Goldberger:1999uk} proposed 
a mechanism  for radius stabilization by showing that a bulk scalar field propagating in the
warped geometry can generate a potential for this radion field and in
the same process, dynamically generate a vacuum expectation value (vev) for
$\varphi(x)$ required to stabilize the radius to the constant value
needed to address the hierarchy of the electroweak (EW) scale and the Planck scale.
Its mass, however, can be much lighter than those of the massive
gravitons \cite{Davoudiasl:1999jd}.  The radion coupling to SM fields
is governed by its vev,  $\Lambda_\varphi
(\simeq$ TeV) and the trace of the energy-momentum tensor ($T^\mu_\mu$)
\cite{Goldberger:1999un}.  At LEP, such a light radion could have been produced
via $e^{+}e^{-}\rightarrow Z \varphi$. The production mode in this channel is however found to be suppressed for $\Lambda_{\varphi} > 1.0 ~\rm TeV$ and hence, 
a radion as light as 50 -- 100~GeV with 
$\Lambda_\varphi \simeq 2 - 3$~TeV, is still allowed by LEP data as well as 
by LHC searches \cite{Mahanta:2000ci,Barate:2003sz,Aad:2014ioa}.
The radion can in principle mix with the
Higgs boson through terms consistent with general covariance.  
The phenomenology of
such a mixed state has been considered in detail in the literature
\cite{Giudice:2000av, Csaki:2000zn, Csaki:1999mp, Dominici:2002jv,
  Csaki:2007ns, Gunion:2003px, Chaichian:2001rq, Rizzo:2002bt,
  Toharia:2008tm, Battaglia:2003gb} and more recently has been
re-investigated \cite{Desai:2013pga, deSandes:2011zs, Kubota:2012in, 
Jung:2014zga, Geller:2013cfa,
  Cox:2013rva, Kubota:2014mma} in light of the discovery of the $\sim$
125~GeV scalar resonance at LHC.  Similarly, the phenomenology of the
simpler scenario of an unmixed radion, too, has been studied quite
thoroughly \cite{Ohno:2013sc, Goncalves:2010dw, Barger:2011hu,
  Cheung:2011nv, Davoudiasl:2012xd, Chacko:2012vm, chacko}. In this work we
restrict ourselves to the unmixed scenario such that the scalar
resonance observed at LHC is a pure SM Higgs boson ($h$). We
concentrate on identifying the most promising signals for an unmixed
light radion ($m_{\varphi} < m_h$), which could provide the
first observable signals for models of extra spatial dimensions with
warped geometry. Our results can be very easily generalized to the
mixed scenario as well,  and are also applicable to extensions of the RS model where the SM fields propagate in the bulk. We focus primarily on the 
following interesting highlights of a light radion signal at the LHC:
\begin{itemize}
\item {An unmixed radion lighter than the 125~GeV Higgs can have
    appreciable production cross section for allowed values of the vev
    ($2 ~{\rm TeV} < \Lambda_\varphi < 3$~TeV), primarily through
    gluon fusion.  A factor that contributes to this, namely, the
    trace anomaly contribution, boosts the loop induced decay
    modes of the radion into a pair of massless gauge bosons.  This
    can partially compensate for the $\Lambda_\varphi$ suppressions in
    its couplings to SM particles. }
\item {A light radion with mass below 100~GeV is not ruled out by any
    experiments \cite{Cho:2013mva}.  We show that the channel ($\gamma\gamma$) which
    helped discover the SM Higgs with the maximum significance
    would also be the most promising channel for such a light radion
    at the LHC.}
\item {The
radion loop-induced decay mode
($\gamma\gamma$) also acquires an enhancement from the trace anomaly
(which interferes constructively with the dominant $W$ boson mediated
loop amplitude) and yields a reasonably healthy, {\it  albeit} small diphoton 
branching rate for radion masses below 120~GeV.}
\end {itemize}

One must note that the radion signal depends crucially on the value of
$\Lambda_\varphi$ which suppresses the effective coupling of the
radion to SM fields as the couplings are inversely proportional to the
value of $\Lambda_\varphi$. Current constraints on the KK
excitations of the spin-2 graviton already put a lower bound on the
value of the $\Lambda_\varphi$ \cite{Aad:2014cka}.  

For a radion of mass $\simeq$ 100~GeV and lower, the dominant
decay modes are gluon-gluon and $b\bar{b}$, while the branching ratios
into $WW^{*}/ZZ^{*}$ are suppressed. The signal arising from
$b\bar{b}$ and $gg$ are beset with large QCD backgrounds, even if we
consider various associated production channels. Thus, with the
enhanced $gg$ fusion as the production mode, $\varphi \to
\gamma\gamma$ becomes the best channel for observing the
light radion at the LHC.  Since a peak in the diphoton invariant mass is a rather spectacular signal of new physics, 
the refinement of techniques to isolate two photons can be helpful in a more general context as well. 

With the impressive performance of the
electromagnetic calorimeter at the CMS and ATLAS experiments, and
optimized event selection criteria for the diphoton signal, we have
been able to observe the SM Higgs boson with large significance,  even
with nominal luminosities available at the 7 and 8 TeV runs. We are
about to enter a regime of higher intensity running of the LHC with
roughly double the center of mass energies. In view of this, the
prospects of observing a light radion in the same mode are good. We demonstrate this with a detailed analysis of the
radion signal and the SM background in the $p p \to \gamma \gamma + X$
events at the 14 TeV run of LHC.

The SM backgrounds for these events are of course formidable.  As for the case of Higgs signals, 
the $\gamma\gamma$ final state has
backgrounds from not only prompt photon pairs, but also $\gamma j$ and
$jj$ production. Of these, the $\gamma j$ background can be
substantial, especially for low diphoton invariant mass. We followed 
the cuts commonly used by ATLAS and CMS for reducing these backgrounds without 
compromising too much on the signal rates \cite{TDR,Aad:2014ioa}.

Our paper is organized as follows. In Section \ref{sec:model} we
describe briefly the RS model with an unmixed radion.  In Section \ref{sec:analysis} we present our
analysis and results for observing the radion in the diphoton channel. We finally
summarize and conclude in Section \ref{sec:conclusions}.  Additional
formulas for production and decay of the radion are provided in the
Appendices (\ref{appendix}).

\section{The radion in models with a warped extra dimension}
\label{sec:model}
In the original version of the Randall-Sundrum model, there is an
extra space-like dimension, namely, y=$r_c\phi$, which is $S^1/Z_2$
orbifolded. Two 3-branes with tensions of opposite signs are
present at the orbifold fixed points $\phi = 0$ and $\phi =
\pi$. Gravity propagates in the bulk and it mainly peaks at the first
brane ($\phi = 0$), called the hidden brane, whereas all other SM
fields propagate on the second brane ($\phi = \pi$), called the
visible brane.  The resulting non-factorisable 5-dimensional metric
depends on the radius of compactification $(r_c)$ of the additional
dimension
\begin{equation}
 ds^2 = e^{-2 k r_c \phi}\eta_{\mu\nu} dx^{\mu} dx^{\nu} + r_c^2 d\phi^2.
\end{equation}
The Planck mass associated with the 4-dimensional  space-time  $(M_{Pl})$ is of the same order of magnitude as the
5-dimensional space-time Planck mass $(M)$. They are related by
\begin{equation}
 M_{Pl}^2 = \frac{M^3}{k}(1 - e^{-2kr_c \phi}).
\end{equation}
A field that propagates on the visible brane in the 5-dimensional
theory carrying a mass parameter $m_{0}$ generates a physical mass $m =
m_{0} e^{- k \pi r_{c}}$ in the 4-dimensional effective theory.  For
the value of $kr_c$ $\simeq$ 12, the Planck scale is reduced to the weak
scale, thus solving the hierarchy problem.

The above metric allows two types of massless excitation. The first one
is the fluctuation of the flat background metric that generates
a bulk graviton. The second conceivable fluctuation is that of the 
compactification radius $r_c$, which can be expressed as $T(x)$,
where $T$ is a modulus field.

The Kaluza-Klein (KK) decomposition of the bulk graviton on
the visible brane generates a discrete tower of states, with the zero
mode as the massless graviton mode. The mass of the n-th KK mode of
the graviton is given by
\begin{equation}
 m_{n} = k x_{n} e^{-k r_c \pi},
\end{equation}
where $x_{n}$ is the n-th root of $J_{1}$, the Bessel function of order 1.
 
The massless mode of the graviton couples to matter with a strength
suppressed by the Planck mass.  The corresponding couplings of the
massive KK modes are suppressed at the TeV scale, with an effective
coupling given by $k/\bar{M}_{Pl}$, where $\bar{M}_{Pl}$ is the
reduced Planck mass. The KK excitations of the graviton can be directly probed at 
the LHC and recent experimental limits from available LHC data rule out the possibility of a mass below 2.67 TeV for the 1-st KK mode graviton with $k/\bar{M_{Pl}}= 0.1$  \cite{Aad:2014cka}.

However, there is one more new physics component of the RS scenario.
The radius $r_c$ of the compact dimension seems to be frozen {\it ad-hoc} at 
the requisite value for solving the hierarchy problem. This arbitrariness
is removed if, as stated earlier, $r_c$ can be construed as the vev of a modulus
field $T(x)$ which quantifies the fluctuation about the stabilized radius.
With this, the metric becomes
\begin{equation}
 ds^2 = e^{-2 k T(x) \varphi}g_{\mu\nu} (x) dx^{\mu} dx^{\nu} + T^2(x) d^2 \varphi .
\end{equation} 
After Kaluza-Klein reduction of the 5-dimensional action and after 
integrating out the additional coordinate, the $T(x)$ dependent part of the action is
\begin{equation}
\label{eqn:action_4d}
 S = 2\frac{M^3}{k} \int d^4x \sqrt{-g(x)} R (1 - e^{-2k T(x) \pi}) + \frac{3 M^3} {k} \int d^4x \sqrt{-g(x)} \partial_{\mu} (e^{-k \pi T(x)}) \partial^{\mu} (e^{-k \pi T(x)}).
\end{equation}
Defining $\varphi(x)$ = $\Lambda_{\varphi} e^{-k [T(x) - r_c] \pi}$ with 
$\Lambda_{\varphi}$ = $ \sqrt{\frac{6 M^3} {k}}e^{-k r_c \pi} $, Eq. [\ref{eqn:action_4d}] becomes

\begin{equation}
\label{eqn:action_radion}
 S = \frac{2M^3}{k} \int d^4 x \sqrt{-g} (1 - (\frac{\varphi}{f})^2) R + \frac{1}{2} \int d^4 x \sqrt{-g} \partial_{\mu} \varphi \partial^{\mu} \varphi .
\end{equation}
This $\varphi(x)$ field is known as the radion field. However, at this point 
there is no mechanism of stabilizing the
radion field such that $T(x)$ acquires its desired vev $r_c$, since 
$\varphi$ is {\it prima facie} massless. This stabilization
is implemented through the Goldberger-Wise mechanism where an additional 
bulk scalar field
is introduced, which develops an effective 4-dimensional
potential on the brane. This potential generates the mass 
as well as the vev of the radion. The
parameters of the potential have to be such that it attains
its minima for $kr_c = 12$.  The mass of the radion, essentially a
free parameter, can be smaller than the TeV scale, even when the massive
graviton modes are much heavier. The principle of general covariance
allows the radion to couple with matter through the trace of the 
energy-momentum tensor. Its interaction with the SM particles is given by
 \begin{equation}
{\cal{L}}_{int} = T^{\mu}_{\mu} \frac{\varphi}{\Lambda_{\varphi}},
 \end{equation}
where $T^{\mu}_{\mu}$ is the trace of energy momentum tensor $T_{\mu \nu}$.
Thus, the interaction of the radion with the massive SM particles is given by 
\begin{equation}
 {\cal{L}}_{1} = \frac{\varphi}{\Lambda_{\varphi}}(\sum_{f} m_{f} f \bar{f} - 2m_{W}^2 W_{\mu}^{+} W^{\mu -} - m_{Z}^2 Z_{\mu} Z^{\mu} + (2 m_{h} h^2 - \partial_{\mu}h \partial^{\mu}h)).
 \label{eq:L1}
\end{equation}
The mass ($m_\varphi$) and vev ($\Lambda_\varphi$) of the radion determine
its phenomenology, similarly to the case of the SM Higgs. 
The mass of the first KK mode of graviton $m_{1}$,  $k/\bar{M}_{Pl}$ and 
the $\Lambda_{\varphi}$ are related by
\begin{equation}
\frac{k}{\bar{M}_{Pl}} = \frac{\sqrt{6}m_{1}}{\Lambda_{\varphi} x_{1}}~~~{\rm with}~x_{1}~=~3.83.
\end{equation}
To suppress higher curvature terms, $\frac{k}{\bar{M}_{Pl}}$ should not be greater than 1.
Thus, the absence of the first KK mode of graviton at the LHC till 2.67 TeV implies
a lower limit of about 1.8 TeV on the radion vev \cite{Aad:2014cka,Grzadkowski:2012ng,Cho:2013mva}.

The effective couplings of $\varphi$ with gluon and photon
pairs are slightly different and have two components.  The first one, 
just like for the SM Higgs, comes from the amplitude of the one-loop
diagrams dominantly involving the top quark, and the $W$ boson for the photon. The second
contribution arises from the trace anomaly for the massless gauge
field. Thus, the interaction of the radion with a gluon pair is given
by
\begin{equation}
 {\cal{L}}_{2} = \frac{\alpha_{s}}{16\pi}G_{\mu \nu}G^{\mu \nu}[2 b_3 - F_{1/2}(\tau_t)]\frac{\varphi}{\Lambda_{\varphi}},
  \label{eq:L2}
\end{equation}
where $b_3 = 7$ is the QCD $\beta$ function.
The effective diphoton interaction of the radion is similarly  given by 
 \begin{equation}
  {\cal{L}}_{3} = \frac{\alpha_{EM}}{8\pi}F_{\mu \nu}F^{\mu \nu}[(b_2 + b_Y) - (F_1(\tau_W) + \frac{4}{3} F_{1/2}(\tau_t))]\frac{\varphi}{\Lambda_{\varphi}},
   \label{eq:L3}
 \end{equation}
where $b_2 = 19/6$ and $b_Y = -41/6$ are the SM $SU(2)$ and $U(1)_{Y}$ $\beta$ functions respectively.
 
In principle, the radion can mix with the SM Higgs via general covariant terms,
which trigger a kinetic mixing. The coefficient of this mixing term can affect
the phenomenologies of both the radion and the Higgs field. As has been stated in
the introduction, our purpose here is to find out signals of a light radion,
for which such mixing is neglected in the first approximation.

\section{Analysis of the radion in the two-photon channel}
\label{sec:analysis}

\subsection{Radion production and decay at the LHC}

At hadron colliders, the radion can be produced via gluon fusion or
through $W$ or $Z$ fusion, and can also have associated production
modes with $W,Z$ bosons and $t\bar{t}$.  The first of the
aforementioned production modes,  receives a sizable boost from
trace anomaly. The radion can also be produced in association with a $W$ or $Z$ boson.
The radion produced in association with a gauge boson can decay to 
$b \bar{b}$ with sizable cross section. The final state will be either
dilepton plus two b-jets or single lepton plus two b-jets.  But the
associated production channel is not of much use, due to its 
suppression by $\Lambda_{\varphi}$, in contrast to the gluon-fusion
channel where the trace anomaly term at least partially compensates
with an enhancement.  We analyzed the 
final states for such a signal and found that the SM dilepton background and 
single lepton background overwhelms the signal and is roughly three to four orders 
of magnitude higher than the signal. Another possibility is the production of 
the radion via vector boson fusion and its subsequent decay to $b\bar{b}$. Here too,
the suppression in couplings by the radion vev is a problem; and on the whole,
the $2j + b\bar b$ SM background is also found to be larger than the signal by
four to five orders of magnitude \cite{CMS:vbf}. The most promising production channel 
thus remains the gluon fusion.

\begin{figure}[ht]
 \centering
   \includegraphics[width=0.5\linewidth, height=0.35\textheight]{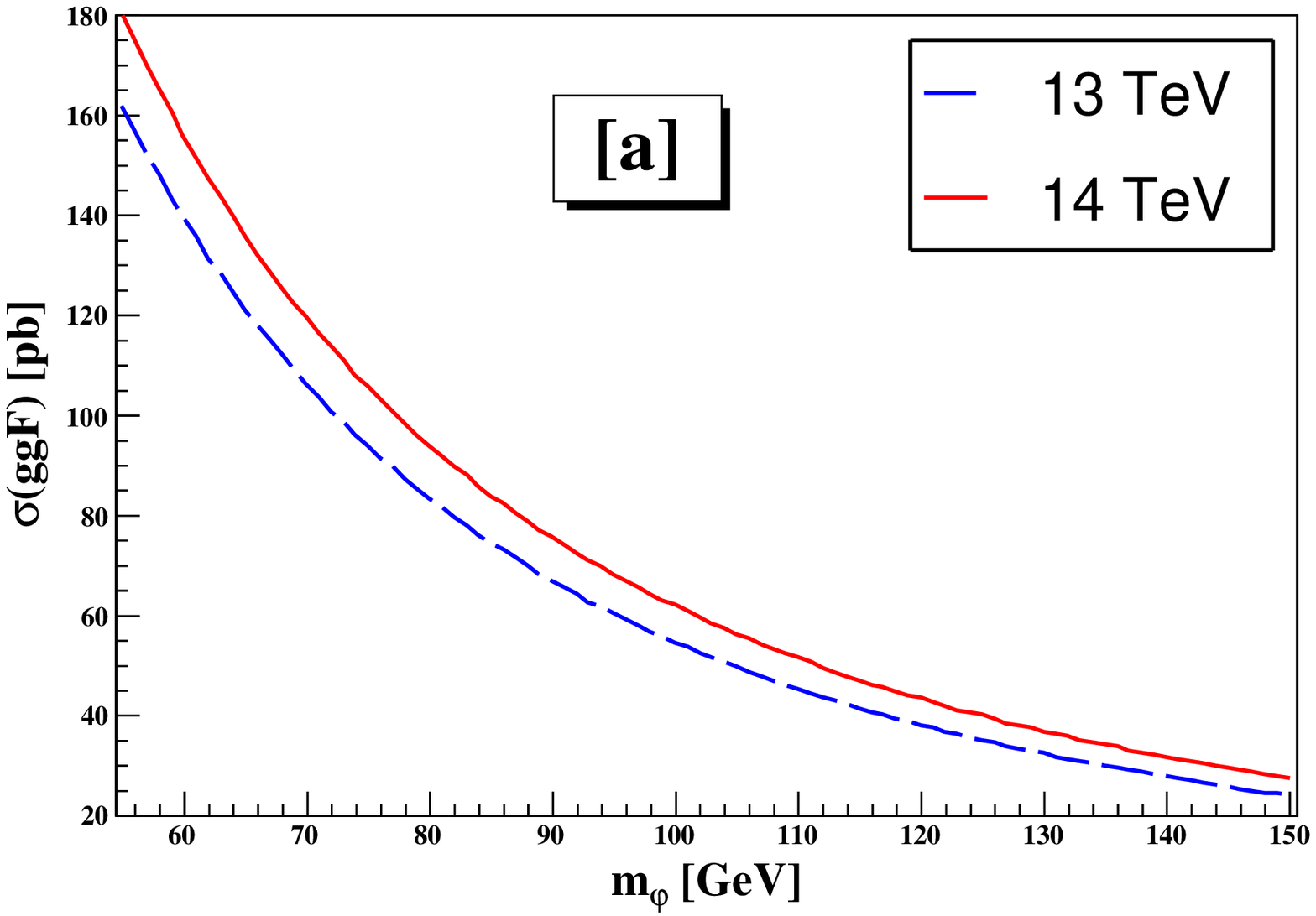}
\hskip 1in
  \includegraphics[width=0.5\linewidth, height=0.35\textheight]{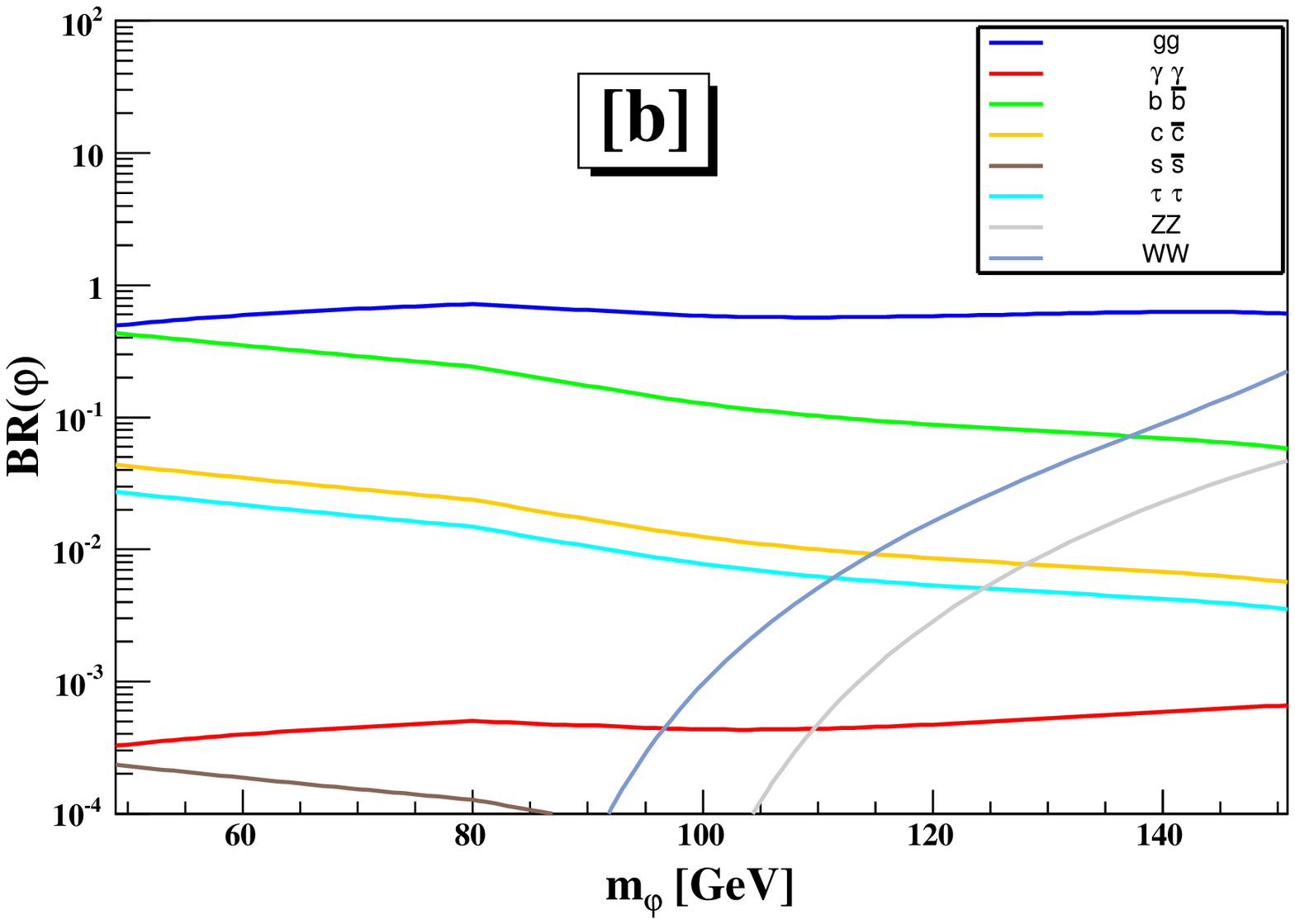}
    \caption{\footnotesize{(a) Production cross section of radion via gluon fusion versus $m_{\varphi}$ for 13 TeV and 14 TeV CM energies at the LHC. (b) Branching ratios for the radion decay modes as functions of its mass $m_{\varphi}$.}}
  \label{fig:decay}
\end{figure}

The production cross section of the radion in gluon fusion channel at the LHC is illustrated in 
Fig \ref{fig:decay}(a) for 13 TeV and 14 TeV center of mass energies.  Since the cross-sections 
are of comparable magnitudes, we present the rest of our results for 14 TeV, with the understanding 
that the predictions are generally valid if a part of the LHC run is at 13 TeV center of mass 
energy.

We used a radion vev,  $\Lambda_{\varphi} =2$ TeV in most of our subsequent analysis. 
The cross-section corresponding to any other $\Lambda_{\varphi}$ can be 
obtained by simple scaling. The branching ratios of the radion to all possible final states
are shown in Fig \ref{fig:decay}(b).  Note that the different branching ratios of the radion decay are independent 
of $\Lambda_{\varphi}$, since all interactions of the radion with SM particles is inversely proportional
to it, including the radion width. 

As seen from Fig \ref{fig:decay}(b), when the mass of the
radion is less than 100~GeV, it decays dominantly into two gluons.
However the two gluon final state gets swamped by the large QCD
background at the LHC, making it a very difficult channel to observe
any signal for a light radion. 

This leaves  two potential channels in which a light radion can
be probed, namely, $\gamma\gamma$ and $\tau^{+}\tau^{-}$. From the 
experience with the Higgs boson,  various subtleties involved
in the analysis of a $\tau^{+}\tau^{-}$ final state makes it more
suitable as a  channel which will confirm the presence of the radion, rather than one used for
discovery. Furthermore, a light radion produces relatively
softer $\tau$'s, which can stand in the way of efficient
identification.  The diphoton final state, on the other hand, is more
spectacular in terms of reconstruction, in spite of the low branching
ratio.  Thus the diphoton channel,
when it comes to uncovering a radion in the mass range 60 - 110~GeV, remains the most 
promising, and which we analyze next.

\subsection{The diphoton channel:  signal and backgrounds}
As stated, the diphoton channel for the radion is one with very
high sensitivity, and should be given priority in the explorations at the 14~TeV run of the
LHC. In our study we have varied the mass of the radion from 60~GeV to 110~GeV. The status of a
heavier radion can be surmised from the 8 TeV run itself, for example,
from reference \cite{Ohno:2013sc,Desai:2013pga} in the zero-mixing limit. 
 The diphoton signal for a radion of mass $m_{\varphi} > 100~\rm GeV$ has 
also been considered in \cite{Davoudiasl:2010fb}. However, that analysis is based on a model with gauge fields
in the bulk, where the diphoton rate receives an enhancement\footnote{Other mechanisms leading to enhancement in the diphoton channel also exist  \cite{chacko}.}. Our study addresses a situation where (a) such enhancement
is absent and (b) the radion is lighter than 100~GeV.  On both counts, overcoming the backgrounds thus becomes a tougher challenge for us.

Two isolated photons in the final state can be mimicked by many SM 
processes. We classify the processes into two categories, reducible and irreducible.
\begin{itemize}
\item The irreducible background consists of two prompt photons in the
  final state. It originates from the tree level production via
  $q\bar{q}$ annihilation (Born process) as well as from the one-loop
  process (box diagram) in gluon fusion with quarks running in the
  loop. The contribution from the latter is comparable to that from
  the Born level process because of the high gluon flux at low-$x$,
 where $x$ represents the energy fraction of the colliding proton energy 
 carried by the partons.
  These photons are as isolated as those arising from radion decay.
  Such isolated photon pairs constitute an irreducible background
  to the signal in any search window for a mass peak \cite{Binoth:1999qq}.

\item The dominant reducible background arises from a prompt photon
  along with a jet.  A $\pi^0$, a $\rho$ or an $\eta$ decays into two
  collimated photons that are identified as a single electromagnetic
  cluster in the detector.  This causes the misidentification of jets
  as hard isolated photons. Although the probability of this
  misidentification in a particular event is small, the sheer volume
  of the $\gamma j$ cross-section turns it into a serious background.
  We suggest ways of reducing this kind of
  background in the subsequent analysis.

\item Similarly, as above, two jets can be misidentified as a pair of isolated
  photons. The double misidentification probability, however, is small, and the 
  dijet background is not significant in the present analysis.

\item The Drell-Yan production of $e^{+}e^{-}$ can also mimic
  diphotons, if the $e^{\pm}$ tracks are not correctly reconstructed by the
  inner tracking chamber.  We  convolute the
  Drell-Yan background with a typical inefficiency of 5\% for the
  track detector at the LHC \cite{pc1}.
\end{itemize}

\subsection{Signal versus Background: The Phoenix-effect}
The signal events are generated in {\tt MADGRAPH 5} \cite{madgraph}, where the interaction
vertices of the radion are included using the {\tt FeynRules} \cite{feynrules} package. We have
used {\tt PYTHIA 8} \cite{pythia} for showering and hadronization of the signal events as
well as for generating background events. We adopted CTEQ6l1 \cite{Pumplin:2002vw} as our
parton density function (PDF). The renormalization and factorization
scales are kept at the default value of {\tt PYTHIA 8}. To obtain sufficient
statistics for the signal as well as for the background events, we 
divided our whole analysis into different phase space regions
distinguished by the value of the radion mass. For this purpose, we 
designated different region of $\rm \hat{m}$  (the invariant mass of the
  outgoing partons), for different mass values of the radion:
  \begin{itemize}
   \item For $ \rm m_{\varphi}$ = 60 GeV: 45 GeV $\leq \rm \hat{m} \leq$ 75 GeV;
\item For $ \rm m_{\varphi}$ = 70 GeV: 55 GeV $\leq \rm \hat{m} \leq$ 85 GeV;
\item For $ \rm m_{\varphi}$ = 80 GeV: 65 GeV $\leq \rm \hat{m} \leq$ 95 GeV;
\item For $ \rm m_{\varphi}$ = 90 GeV: 75 GeV $\leq \rm \hat{m} \leq$ 105 GeV;
\item For $ \rm m_{\varphi}$ = 100 GeV: 85 GeV $\leq \rm \hat{m} \leq$ 115 GeV;
\item For $ \rm m_{\varphi}$ = 110 GeV: 95 GeV $\leq \rm \hat{m} \leq$ 125 GeV.
  \end{itemize}

For realistic background estimations, we implemented an algorithm
at the generator level, which approximates the clustering procedure in 
a typical  electromagnetic calorimeter (ECAL). 
Specifically, we used the dimension of an ECAL crystal of the
Compact Muon Solenoid (CMS) detector. The ECAL at the CMS is made up of 
Lead Tungstate ($\rm PbWO_{4}$) crystals. A single crystal of the ECAL covers
$0.0175\times0.0175$ in the $\eta-\phi$ plane. The electromagnetic shower from an unconverted photon
is contained within a $5\times5$ crystal matrix around the seed crystal ({\it i.e.}, the
one hit by the photon). In case of a converted photon, the typical region
 of energy deposit is wider. In order to make the analysis robust,
we used a $10\times10$ crystal size for photon reconstruction,
 equal to $\rm \triangle R = 0.09$
(where $\rm \triangle R = \sqrt{\ {\triangle \eta}^2 + {\triangle
 \phi}^2}$) in the $\eta-\phi$ plane of the CMS detector.  The
momentum of the photon candidate is defined as the vector sum of the
photon and electron momenta falling within the cone $\rm \triangle R$ = 0.09 around
the seed, which is either a direct photon or an electron.

To account for finite detector resolutions,  we smeared the photon, electron and jet energies  
with Gaussian functions \cite{CMS:2010zta}.  We selected the photon seeds satisfying
$|\eta| < 3.0$. The reconstructed photon candidates are then accepted
if they satisfy the preselection criteria given as
 \begin{itemize}
 \item ${\rm p_{T}^{\gamma,\, leading}} > 15$ GeV and ${\rm p_{T}^{\gamma,\, subleading}} > 10$ GeV;
 \item $ |\eta_{\gamma}| < 2.5 $.
\end{itemize}

The $|\eta|$-interval is reduced further to emulate the inefficient
tracker region. These triggered photon candidates are required to have
minimal hadronic activity.  Jets are reconstructed in our analysis
with $|\eta| < 4.5$ and $\rm {p^{j}_{T}}$  > 10 GeV using an anti-$\rm k_t$
algorithm \cite{Cacciari:2008gp}.  Photons arising from the jets are rejected by demanding that 
the scalar sum of the entire transverse energy within a cone of $
\triangle R = 0.4$  be less than 4 GeV\footnote{This is an 'absolute isolation' criteria. 
One can alternatively require a relative isolation, demanding that the total 
visible $\rm p_{T}$ within $\Delta \rm R = 0.4$ is less than $10 \rm \%$ from that of the 
photon. This raises the statistical significance for lower mass of $m_{\varphi}$.}. Only those
isolated photons which survive the above selection criteria qualify for
our final analysis.

\begin{figure}[ht]
\includegraphics[width=3.15in]{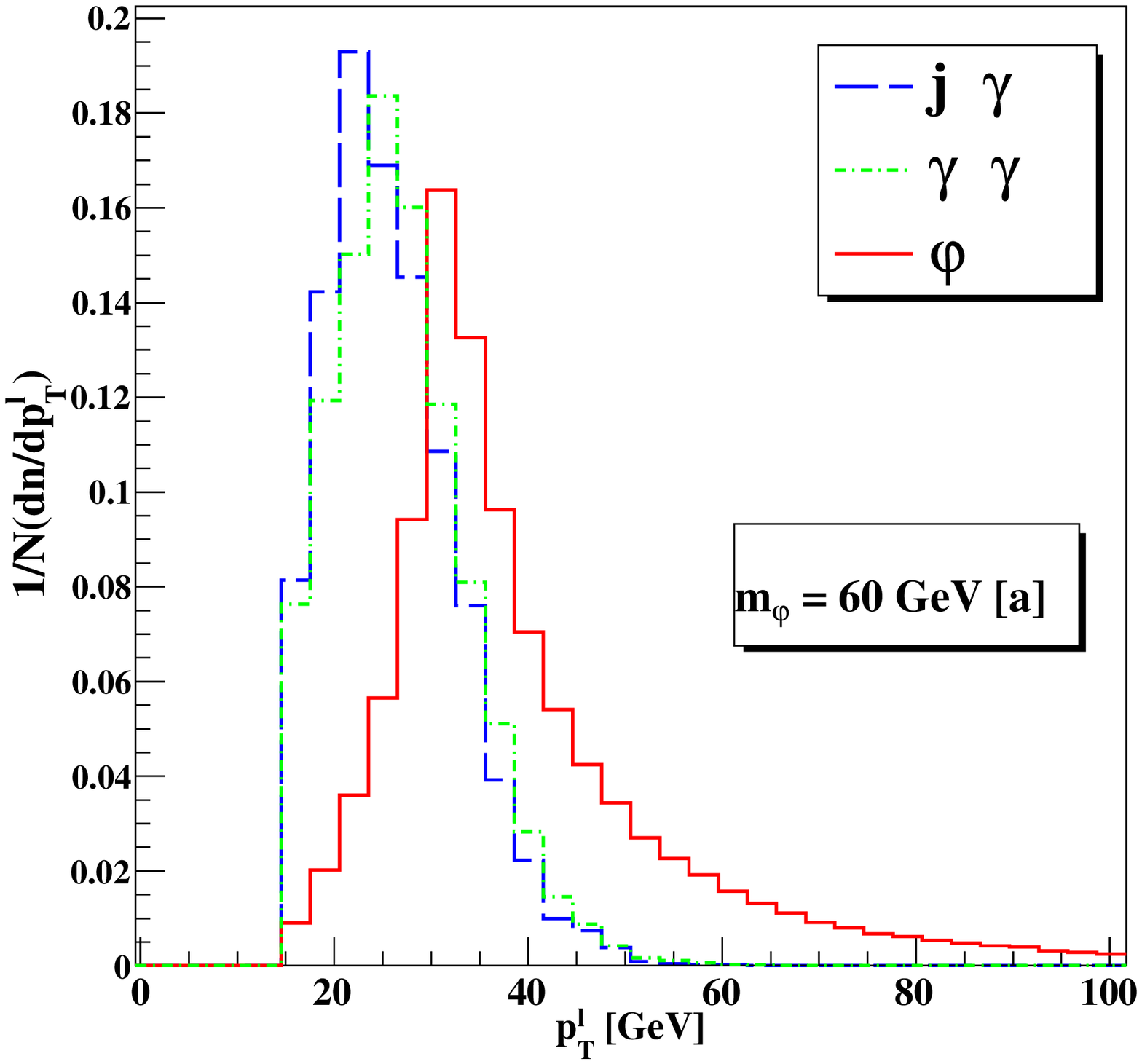}
\includegraphics[width=3.15in]{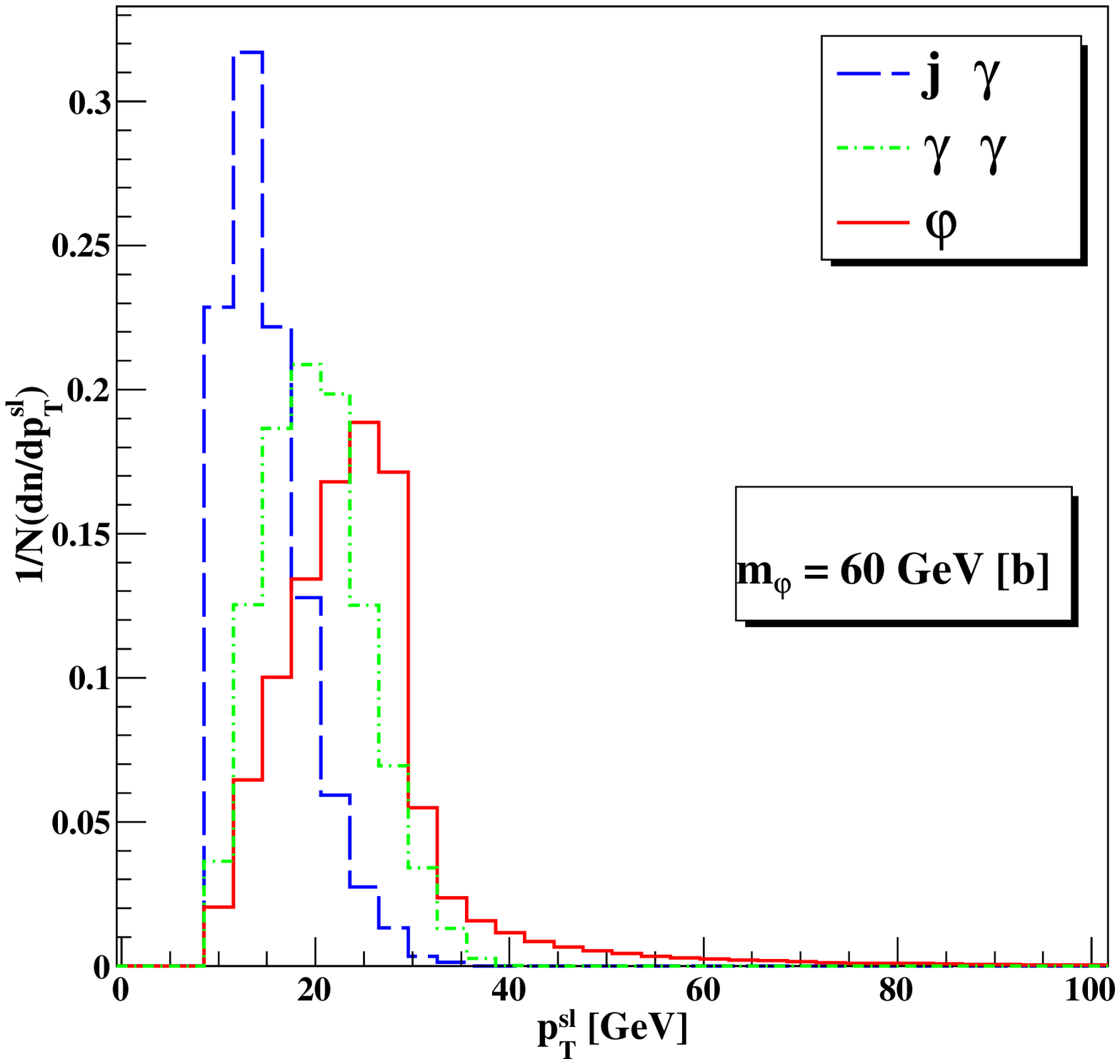}
\includegraphics[width=3.15in]{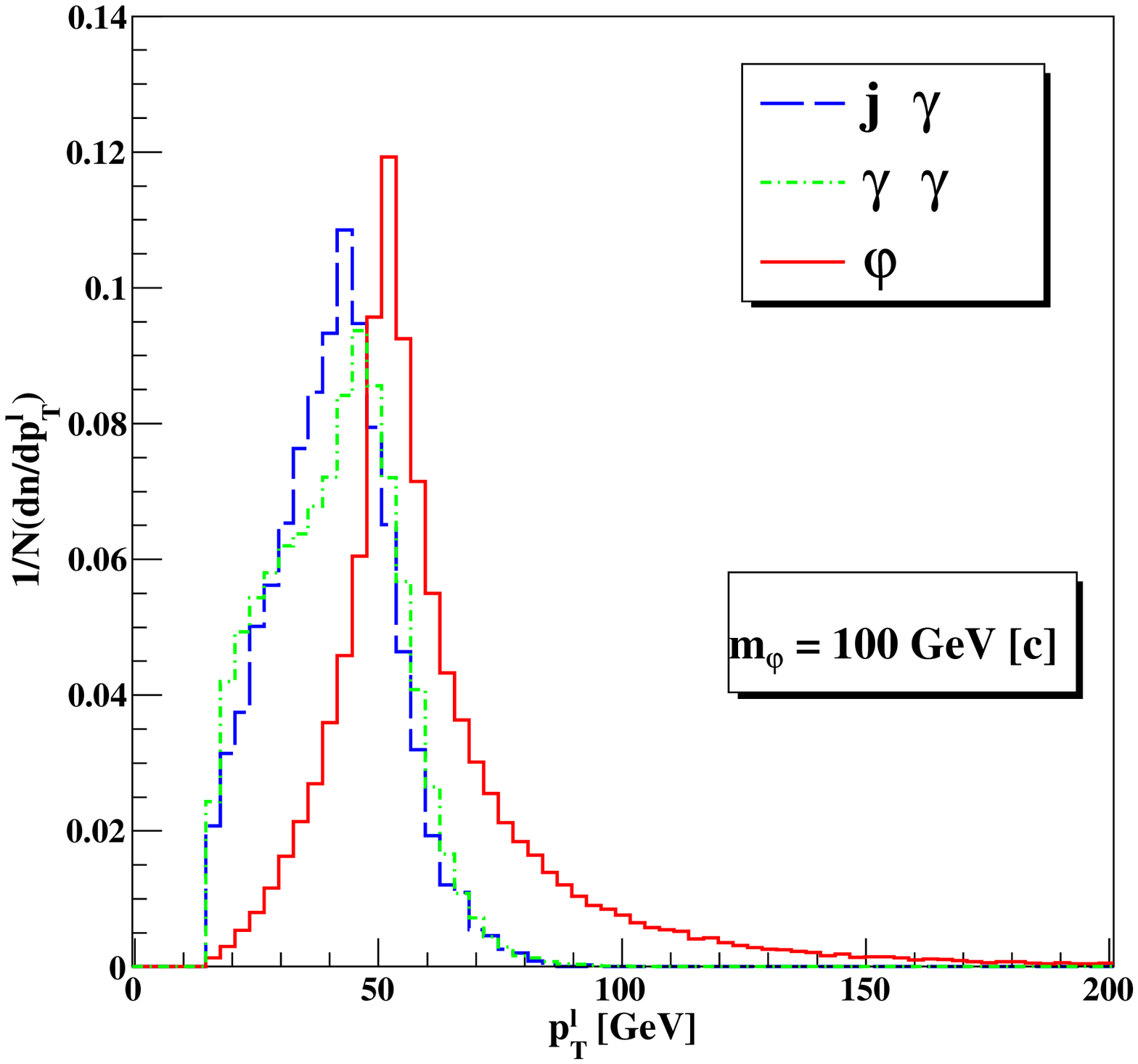}
\includegraphics[width=3.15in]{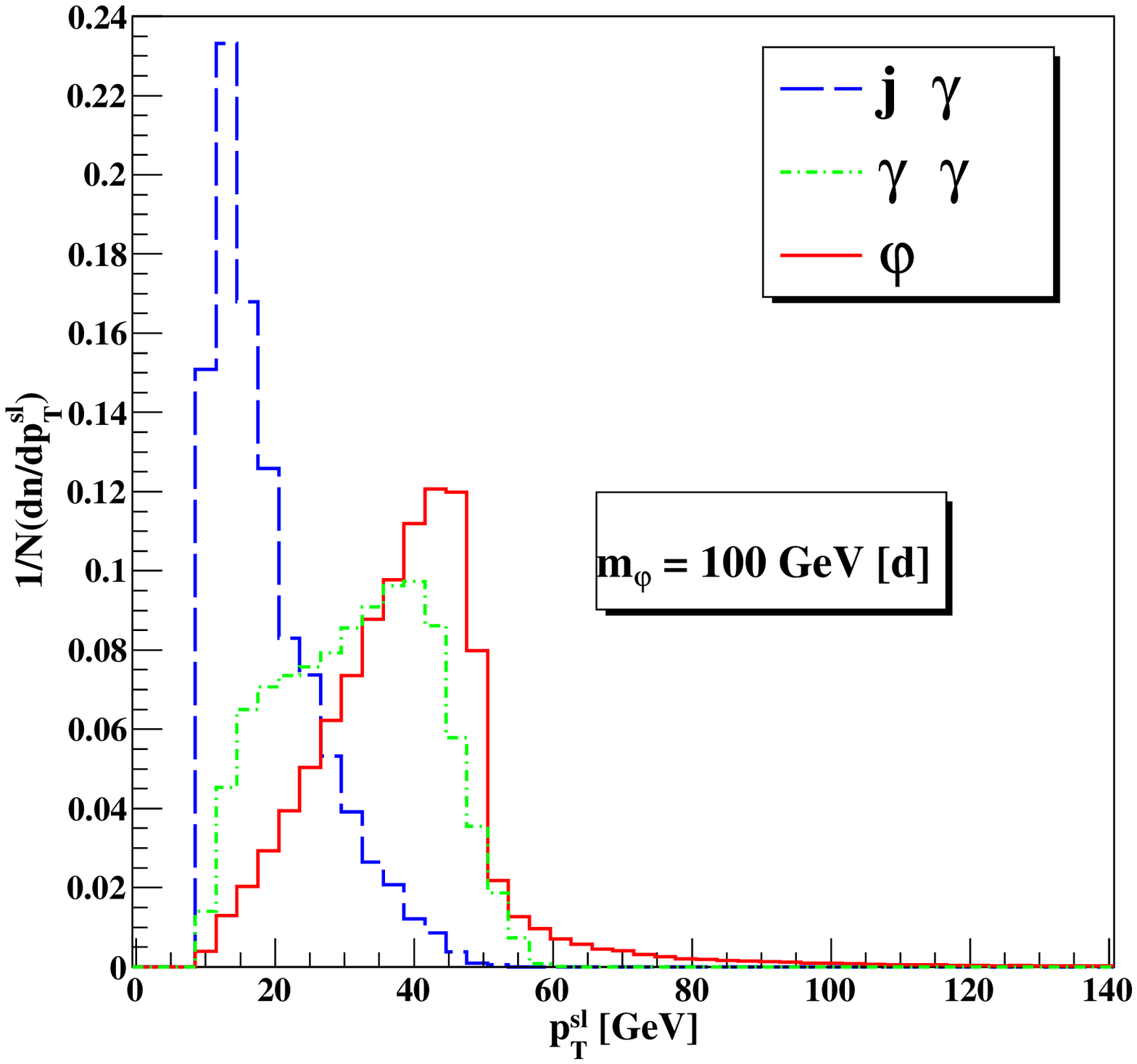}
\caption{Normalized distibution of $\rm p_{T}^{\gamma}$ for two sample masses of radion, diphoton background and signal photon background. 
         (a) Normalized distribution of ${\rm p_{T}^{\gamma,\, leading}}$ for $m_\varphi=60$~GeV;
         (b) Normalized distribution of ${\rm p_{T}^{\gamma,\, subleading}}$ for $m_\varphi=60$~GeV;
         (c) Normalized distribution of ${\rm p_{T}^{\gamma,\, leading}}$ for $m_\varphi=100$~GeV;
         (d) Normalized distribution of ${\rm p_{T}^{\gamma,\, subleading}}$ for $m_\varphi=100$~GeV.}
\label{fig:pT_l}
\end{figure}

The $\rm {p^{\gamma}_{T}}$ distribution for background and signal are plotted in 
Fig. \ref{fig:pT_l}(a), \ref{fig:pT_l}(b) for $m_\varphi=60$ GeV, and in Fig. \ref{fig:pT_l}(c), \ref{fig:pT_l}(d) 
for $m_\varphi=100$~GeV.
 Other kinematic variables, such as angular separations, can be used as good discriminators 
 at the generator level. However, once the detector resolutions are taken into account
the distinct features of these variables are smeared.
We  find that the background coming from a prompt photon and a
jet dominates over the two prompt photon background in the low ($\rm p_{T}^{\gamma}< 35$
GeV)  region. With increasing $\rm
p_{T}^{\gamma}$, the jet-$\gamma$ misidentification rate decreases and
hence the $\gamma j$ background falls gradually. Though the Drell-Yan
background is two orders of magnitude lower than the direct photon
background, it increases near the $Z$ mass pole, and is comparable to
the direct photon backgrounds.  We find that the two-jet
background is negligible, and thus we do not consider it in our
analysis. As seen in Fig. \ref{fig:pT_l}, radion mass-specific $ \rm p_T^{\gamma}$-cuts are effective,
in view of the fact that a heavier radion generally yields harder
photons.  For a heavier radion, the fraction of events with harder
$\rm p_{T}^{\gamma}$ in the signal is large compared to the
background. Thus, it is easier to separate the signal events from the
background by selecting harder photon candidates. 
The mass dependent $\rm p_{T}$ cuts in our analysis are formulated as 
\begin{align}
\rm p_{Tmin}^{leading} = (m_{\varphi}/2 -5.0)~GeV  &;&  \rm p_{Tmin}^{subleading} = (p_{Tmin}^{leading} - 5.0)~GeV . 
\end{align}

 We  finally select only those events that fall within the invariant mass window of
$\pm$3.5~GeV about the radion mass. If we consider the invariant mass
window to be about 5~GeV,  the background rate increases, thus
reducing the signal-to-background significance ($S/\sqrt{B}$).

\begin{figure}[h]
 \centering
   \includegraphics[width=3.2in]{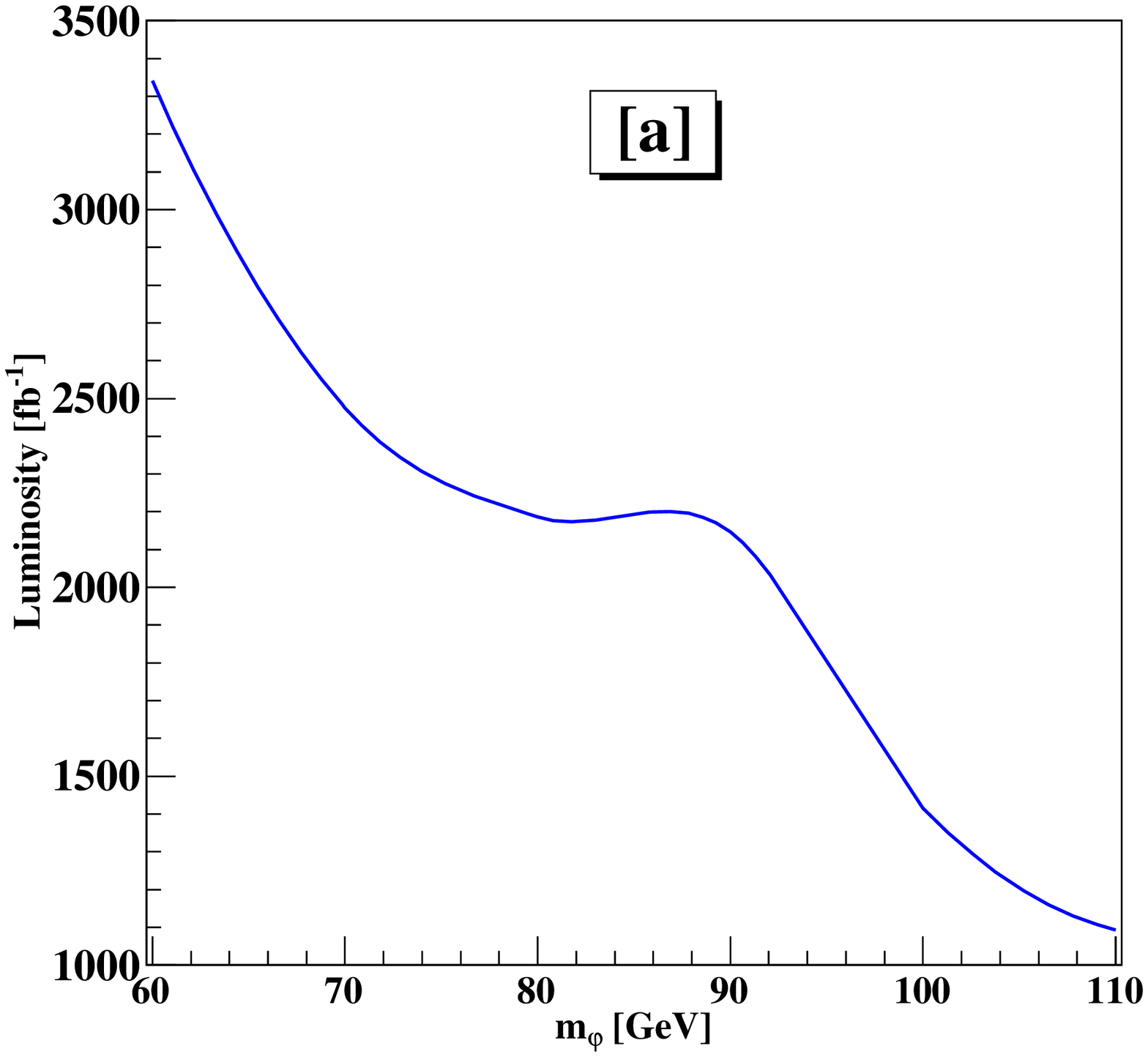}
\vskip 0.1in
  \includegraphics[width=3.2in]{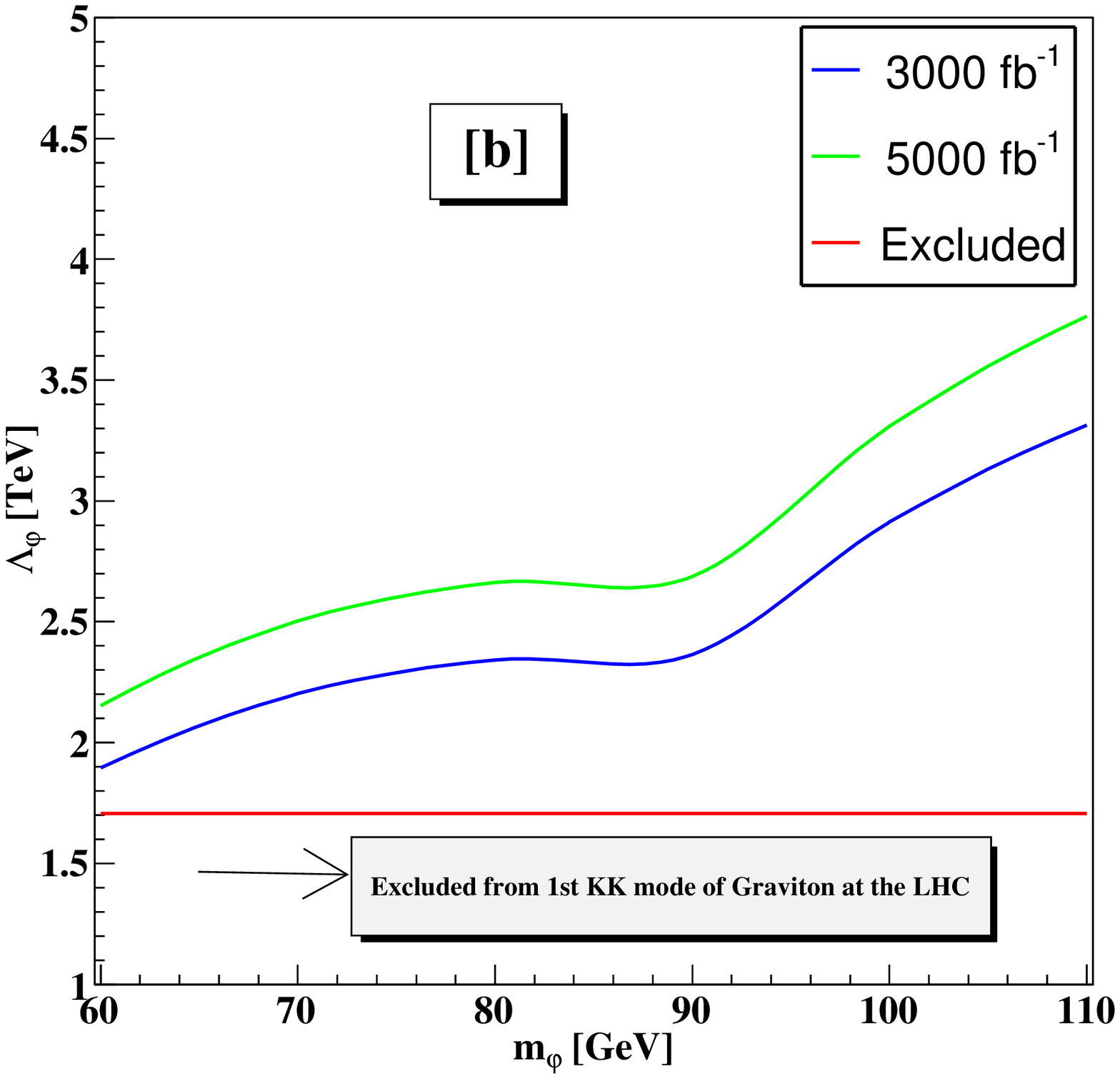}
    \caption{\footnotesize{(a) Luminosity required for 5 $\sigma$ discovery of radion with $m_{\varphi}$ with $\Lambda_{\varphi} = 2 \rm~TeV $. (b) Maximum $\Lambda_{\varphi}$ for a radion to be discovered at 5$\sigma$ with $m_{\varphi}$.}}
  \label{fig:sigma}
\end{figure}

The cut flow for the  signal with 60~GeV and 90~GeV
radion mass and the  corresponding SM background are presented in Table
\ref{table:efficiency}.~The mass dependent cuts along with the final
signal-to-background significance are shown in Table \ref{table:significance}. In Fig. \ref{fig:sigma}(a), we 
plot the integrated luminosity required to achieve 5$\sigma$
significance level for different radion mass. In Fig. \ref{fig:sigma}(b), we also plot the maximum vev of 
the radion that can be probed with 5 $\sigma$ significance level for different mass values of the 
radion with two choices of the integrated luminosity. 
Note that these results do not conflict with the recent ATLAS 
search \cite{Aad:2014ioa} at $\sqrt{s}=8$ TeV and with luminosity ${\cal L}=20.3$ fb$^{-1}$. The 
data rules out signals with $\sigma_{gg} \times BR(\varphi \to \gamma \gamma)$ of 30 fb or more, 
while the signal rate for  $\sqrt{s}=8$ TeV in our scenario is smaller in magnitude.
\begin{table}[ht]
 \begin{center}
  \begin{tabular}{|l|l|l|l|l|l|l|}
   \hline
   {\bf $m_{\varphi}$} &{ Cuts applied} & {\bf $\varphi \rightarrow \gamma \gamma$}& {\bf $\gamma \gamma$} 
     & {$j \gamma$}  & {\bf $e^{+}e^{-}$} & $b_{1} + b_{2} + b_{3}$ \\
   \hline
   { [GeV}]&	       & { S [fb]} & { $b_{1}$ [pb]} & {$b_{2}$ [pb]}& {$b_{3}$ [pb]}& { {B} [pb]} \\
   \hline
   &Initial Signal         & 39.88 & 226.84 & 218109.90 & 133.78& 218470.52 \\ \cline{2-7}
   &Preselection         & 30.80  & 87.88 & 6332.58 & 0.67& 6421.13 \\ \cline{2-7}
   \multirow{1}{*}{60}&Isolation         & 24.51  & 76.76 & 973.20 & 0.55& 1050.51\\	\cline{2-7}
    &${\rm p^{ \gamma,\,l}} > 27$ GeV	&  &  &   &  &  \\
    &$\rm p^{ \gamma,\,sl}_{T} > 22$ GeV  & 14.02 & 19.15 & 49.73 & 0.22 & 69.10\\
    &$ 56.5 < \rm m_{\gamma \gamma} < 63.5$ & 13.98 &6.35 & 22.68  & 0.05 & 29.08\\
    &$\rm {[GeV]}$ &  &  &  &  &  \\	\cline{2-7}
   \hline
   &Initial Signal         & 30.84 & 48.28 & 46788.40 & 1598.90 & 48435.58 \\\cline{2-7}
   &Preselection         & 25.00  & 18.20 & 3198.46 & 10.60&3227.26 \\\cline{2-7}
   \multirow{1}{*}{90}&Isolation         & 19.50  & 15.59 & 309.65 & 8.48& 333.72 \\\cline{2-7}	
    &${\rm p^{ \gamma,\,l}} > 40$ GeV	&  &  &   &  &  \\
    &${\rm p^{ \gamma,\,sl}}> 35$ GeV  & 9.59 & 3.77 & 7.29 & 3.72 & 14.78\\
    &$ 86.5 < \rm m_{\gamma \gamma} < 93.5$ & 9.58 & 1.04 & 2.15  &  2.44 & 5.63 \\
    &$\rm {[GeV]}$ &  &  &  &  &  \\	\cline{2-7}
  \hline
  \end{tabular}
 \end{center}
 \caption{Cut flow table for two different values of radion mass, $m_{\varphi}=60$ GeV and $m_{\varphi}=90$ GeV.}.
\label{table:efficiency}
\end{table}
\begin{table}[ht]
 \begin{center}
  \begin{tabular}{|c|c|c|c|c|c|}
   \hline
   {\bf $m_{\varphi}$} & {\rm $p^{\gamma,\,leading}_{T},~ p^{\gamma,\,subleading}_{T}$} & {\bf $m^{min}_{\gamma \gamma}, ~m^{max}_{\gamma \gamma} $} & { S}& { B}
   & {\bf $\sigma$}     \\
   \hline
   { [GeV] } & { [GeV] } & { [GeV] } & { [fb]} & { [pb]} & $S/\sqrt{B}$\\
   \hline
   60         & 27.0,~22.0 &  56.5,~63.5 & 13.98 & 29.07 & 4.49 \\
   \hline
   70         & 30.0,~25.0  & 66.5,~73.5 & 13.78 & 15.50 & 6.06 \\
   \hline
   80         & 35.0,~30.0  & 76.5,~83.5 & 11.42 & 8.31 & 6.86 \\
   \hline
   90 & 40.0,~35.0 & 86.5,~93.5 & 9.58 & 5.63 & 6.99 \\
   \hline	
   100	& 45.0,~40.0 & 96.5,~103.5 & 8.21 & 1.80  & 10.60 \\		
    \hline
   110 & 50.0,~45.0 & 106.5,~113.5 & 7.04  & 0.79 & 13.72 \\		
  \hline
  \end{tabular}
 \end{center}
 \caption{Selection cut, background reduction and significance at $14$
   TeV cm energy and $3000\; {\rm fb}^{-1}$ integrated luminosity for
   different values of radion mass, $m_{\varphi}.$ The signal-to-background significance,
   $\sigma$ is defined by $S/\sqrt{B}$.}
\label{table:significance}
\end{table}

\begin{figure}[h]
 \centering
   \includegraphics[width=0.6\linewidth, height=0.4\textheight]{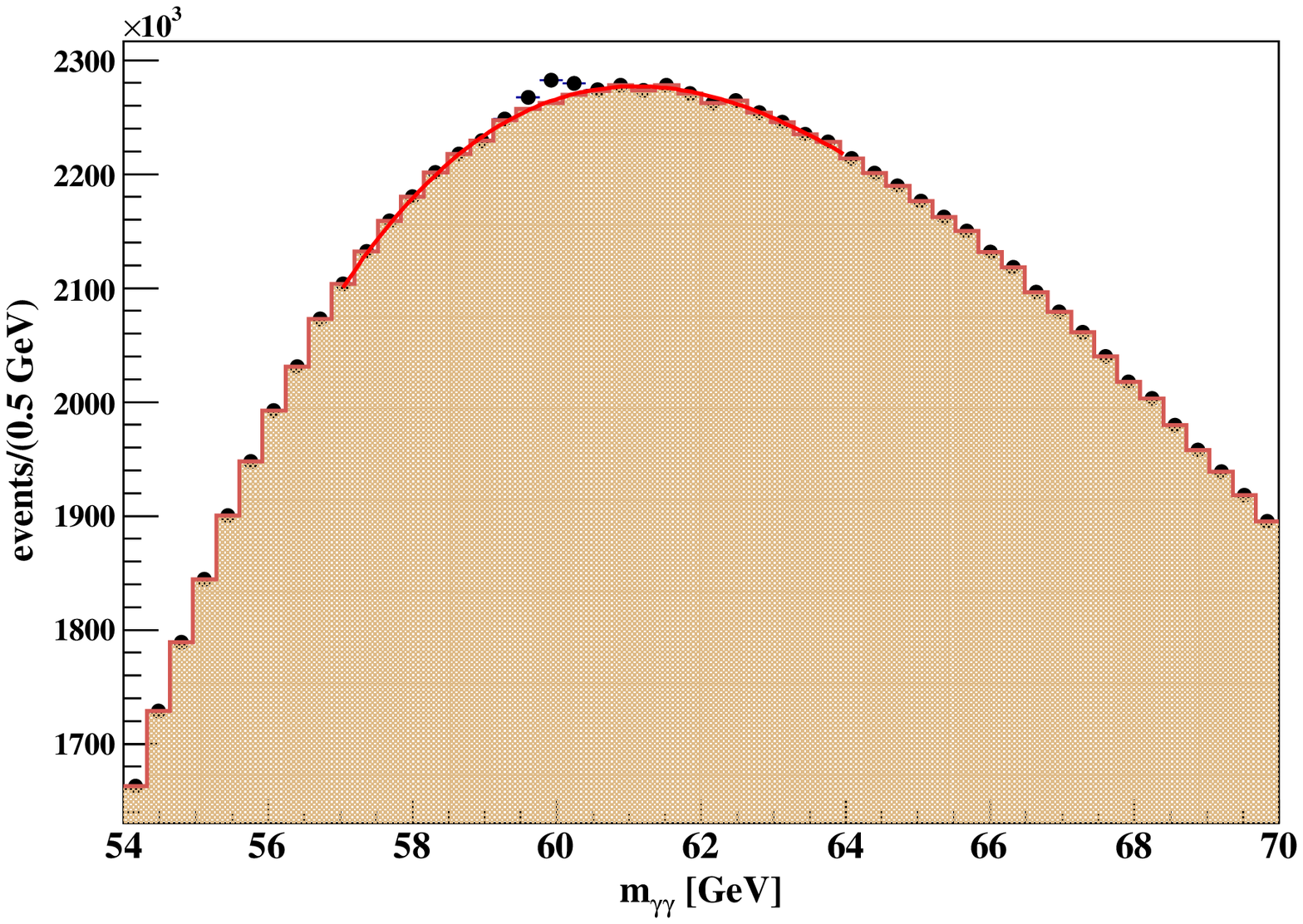}
    \caption{\footnotesize{Invariant mass peak of the signal against the background, for $m_{\varphi}=60~\rm GeV$}.} 
  \label{fig:inv_60}
\end{figure}
Fig. \ref{fig:inv_60} shows the invariant mass peak of the signal against the background, 
for $m_{\varphi} = 60 \rm~GeV$.
For an efficient modeling of the background, a low-luminosity histogram for the background 
has been generated first. Thereafter, a fitting function has been used to improve it, thus yielding the 
background for a luminosity of $3000 \rm~fb^{-1}$.
It should also be noted that the bump corresponding to the signal is sitting on the edge of the rising 
part of the background. This is in contrast with the familiar figure for Higgs reconstruction, where the 
bump is seen against a monotonically falling background profile. This effect is due to the strong 
$\rm p_{T}-\rm cuts$ that we must impose on the photons,  causing an additional background suppression
for low $m_{\gamma\gamma}$ \footnote{ It should be noted that we have assumed perfect 
identification of the vertex from where the photon is coming. In reality, due 
to presence of pileup vertices, photon vertex identification has a finite efficiency, 
which can degrade the mass resolution, and consequently the significance.}.

At this point, we should emphasize that we have carried out our
analysis at the leading order (LO). To estimate how the 
predictions differ when including next-to-leading order (NLO) effects, one notices that the
$K$-factor for the production of an 80 GeV Higgs is approximately 2.0 \cite{pc}. 
For diphotons (including the fragmentation contribution), the same $K$-factor is around 
1.3 \cite{Kdiphoton}. Therefore, the inclusion of the NLO effects will,
if anything, enhance our predicted significance.  We also estimated the effects of varying
the renormalization and factorization scales, which are set to be equal. The results presented here 
are based on using the default value for the renormalization scale ($Q^2$) of the event generator.  
Changing the scale to $Q^2 =  {m_{\gamma\gamma}}^2$ and calculating the uncertainty by varying 
the scale from $Q^2/2$ to $2 Q^2$, the signal as well as the background event rates change by 
about $\pm$ 10\%.

To report the significance of a diphoton mass 
peak we have used a simple $S/\sqrt(B)$ statistic. An alternative analysis using a likelihood ratio is also 
possible \cite{TDR,Cowan:2010,Bhattacharya:2007gs}. While our cut-based analysis is illustrative in nature, 
there is scope for improving the sensitivity of this channel by using more sophisticated techniques. 
if for example, one uses multivariate 
techniques, then the signal significance improves by a factor of 2. Furthermore, on splitting the 
sample in several categories of different purities, one expects an enhancement of about 1.5 times 
in signal significance.

\section{Summary and Conclusions}
\label{sec:conclusions}

While graviton excitations are immediately recognizable signals of warped extra dimensions, 
spectacular as such signals can be,
the limit on the mass of the lowest such excitation is increasing rather rapidly. In view of this it is
important to realize that the radion, connected in a compelling way to the
stabilization of the extra dimension(s), can still be quite light, consistently with
data available so far.  

In this work, we indicated  a method for detecting the signature of a light radion, in
the range 60 - 110~GeV, at the LHC.  After analyzing all production and decay mechanisms, 
the diphoton decay channel following gluon fusion production emerges as
the best and most promising signal.  We thus focused on a pair of photons
reconstructed to a peak at various mass windows, and applied cuts that
can potentially suppress the backgrounds, where the prompt
$\gamma\gamma$ production (at both the Born and box diagram levels)
constitute the irreducible SM backgrounds. Event
selection criteria have been suggested to reduce this as well as the (dominant) 
$\gamma j$ background, where the latter is responsible for producing a fake
photon. After carrying out a detailed study using parametrized simulation and taking into account
all backgrounds, we find that one can separate the signal with a
significance of 5$\sigma$ or more, for an integrated luminosity of
up to 3000 $fb^{-1}$.  In general, less luminosity is required for a 
higher radion mass, as the background falls rapidly
with increasing diphoton invariant mass. The diphoton mode also
avoids any problem near the Z-pole, except of course the possibility
of fakes from electron-positron pairs, which is found to be small.

Notwithstanding the fact that the original RS model  has gone through several extensions 
where SM fields have been allowed to move in the bulk,  radion phenomenology has not become 
markedly different in such extended versions. Thus  our results are valid even in extensions of the 
RS model that allow SM fields in the bulk. 
Moreover, we have studied here the case of the unmixed radion.
 If the radion and the Higgs boson are allowed to mix, under certain circumstances  
this mixing could enhance the mixed radion-Higgs diphoton decay rate. For positive mixing parameter, 
the branching ratio of the light mixed radion (till 150 GeV) 
decaying to diphoton increases and hence can be probed with the diphoton channel 
at the LHC \cite{Toharia:2008tm}. We shall explore this possibility in further studies.  

In an earlier work some of us showed that the LHC data at 8~TeV can constrain the radion 
rather effectively, in a mass range upward of
110~GeV. And now we have found that the range below 110 GeV, all the way down to
60~GeV, is also accessible to probe at the LHC,  for the integrated luminosity 
crossing the attobarn level.


\acknowledgments
We thank Anushree Ghosh, Manoj K. Mandal and V. Ravindran for helpful comments.
The work of UM, BM and SKR was partially supported by funding
available from the Department of Atomic Energy, Government of India,
for the Regional Centre for Accelerator-based Particle Physics
(RECAPP), Harish-Chandra Research Institute.Computational work for this study was par-
tially carried out at the cluster computing facility in the Harish Chandra Research Institute
(http://cluster.hri.res.in). M.F.  is
supported in part by NSERC under grant number SAP105354 and would like to thank Manuel Toharia for illuminating discussions on radions. K.H. acknowledges support from the
Academy of Finland (Project No. 137960). MF, BM and SKR thank the Helsinki Institute of Physics, while  M.F. and K.H. thank  RECAPP for  hospitality at various stages of the project.

 \section{Appendix:  Decay rates of the radion}
 \label{appendix}
 
 \label{subsec:decay}
 
 $\bullet$~~~{ \bf Tree-level decay rates for $\varphi$}
 
 The decay widths of the radion to the SM particles are easily 
calculated from Eqs.~(\ref{eq:L1}, \ref{eq:L2}, \ref{eq:L3}), see also \cite{Cheung:2000rw}: 
\begin{eqnarray}
\Gamma(\varphi\rightarrow f\bar{f}) &=&
\frac{N_c m_f^2 m_\varphi}{8\pi\Lambda_\varphi^2}
                             (1-x_f)^{3/2},\\
\Gamma(\varphi\rightarrow W^+ W^-) &=&
\frac{m_\varphi^3}{16\pi\Lambda_\varphi^2}\sqrt{1-x_W}
                            \Big(1-x_W+\frac{3}{4}x_W^2\Big),\\
\Gamma(\varphi\rightarrow ZZ) &=&
\frac{m_\varphi^3}{32\pi\Lambda_\phi^2}\sqrt{1-x_Z}
                            \Big(1-x_Z+\frac{3}{4}x_Z^2\Big),\\
\Gamma(\varphi\rightarrow hh) &=&
\frac{m_\varphi^3}{32\pi\Lambda_\varphi^2}\sqrt{1-x_h}
                            \Big(1+\frac{1}{2}x_h\Big)^2.
\end{eqnarray}
 The symbol $f$ denotes all
quarks and leptons. The variable $x_i$ 
is defined as $x_i = 4m_i^2/m_\varphi^2\ (i=t,f,W,Z,h)$.  

$\bullet$~~~{ \bf Loop-induced decay rates for $\varphi\to \gamma\gamma,~gg$}

\begin{eqnarray}
\Gamma(\varphi\rightarrow gg) &=&
\frac{\alpha_s^2 m_\varphi^3}{32\pi^3\Lambda_\varphi^2}
          \left| b_{3}+x_t\left\{1+(1-x_t)f(x_t)\right\}\right|^2,
\label{radiongg}
\\
\Gamma(\varphi\rightarrow \gamma\gamma) &=& 
\frac{\alpha_{\rm EM}^2 m_\varphi^3}{256\pi^3\Lambda_\varphi^2}
           \Bigg| b_2 + b_Y - \left\{2+3x_W+3x_W(2-x_W)f(x_W)\right\}
						\nonumber \\
           & & \qquad \qquad \qquad \qquad \qquad
 \qquad +\frac{8}{3}x_t\left\{1+(1-x_t)f(x_t)\right\}\Bigg|^2,
\label{radiongamgam}
\\
\Gamma(\varphi\rightarrow Z\gamma) &=&
\frac{\alpha_{\rm EM}^2 m_\varphi^3}{128\pi^3 s_{\rm w}^2
\Lambda_\varphi^2}\Bigg(1-\frac{m_Z^2}{m_\varphi^2}\Bigg)^3\nonumber\\
& &  \qquad \qquad \qquad 
\times \Bigg| \sum_f N_f \frac{Q_f}{c_{\rm W}} \hat{v}_f \
 A_{1/2}^\varphi(x_f,\lambda_f) +A_1^\varphi(x_W,\lambda_W)\Bigg|^2.
\label{radionZgamma}
\end{eqnarray}
Here, as before $x_i = 4m_i^2/m_\varphi^2\ (i=t,f,W,Z,h)$, and
$\lambda_i = 4m_i^2/m_Z^2\ (i=f,W)$. Here $(b_{3},b_2,b_Y)=(7, 19/6,-41/6)$.
The gauge couplings for QCD and QED are given by $\alpha_s$ and
$\alpha_{\rm EM}$, respectively.
The factor $N_f$ is the number of active quark flavors in the 1-loop
diagrams and
$N_c$ is 3 for quarks and 1 for leptons. 
$Q_f$ and $\hat{v}_f$ denote the electric charge of the
fermion and the reduced vector coupling in the $Zf\bar{f}$ interactions
$\hat{v}_f=2I_f^3-4Q_f s_W^2$, where $I^3_f$ denotes the weak isospin
and $s_W^2\equiv \sin^2{\theta_W},\ c_W^2=1-s_W^2$.

The form factors $A^\varphi_{1/2}(x,\lambda)$ and $A^\varphi_1
(x,\lambda)$ are given by
\begin{align}
A^\varphi_{1/2}(x,\lambda)&=
I_1(x,\lambda)-I_2(x,\lambda)\ ,\\
A^\varphi_1(x,\lambda)    &=
c_W\Bigg\{4\Bigg(3-\frac{s_W^2}{c_W^2}\Bigg)I_2(x,\lambda)
+\Bigg[\Bigg(1+\frac{2}{x}\Bigg)\frac{s_W^2}{c_W^2}-\Bigg(5+\frac{2}{x}
\Bigg)\Bigg]I_1(x,\lambda)
\Bigg\}\ \notag.
\end{align}
The functions $I_1(x,\lambda)$ and $I_2(x,\lambda)$ are
\begin{align}
I_1(x,\lambda) &= \frac{x\lambda}{2(x-\lambda)}
    +\frac{x^2\lambda^2}{2(x-\lambda)^2}[f(x^{-1})-f(\lambda^{-1})]
    +\frac{x^2\lambda}{(x-\lambda)^2}[g(x^{-1})-g(\lambda^{-1})]\ ,
\notag\\
I_2(x,\lambda) &=
 -\frac{x\lambda}{2(x-\lambda)}[f(x^{-1})-f(\lambda^{-1})]\ ,
\label{loop}
\end{align}
where the loop functions $f(x)$ and $g(x)$ in (\ref{radiongg}),
(\ref{radiongamgam}) and (\ref{loop})
are given by
\begin{align}
 f(x) &=
\begin{cases}
    \left\{\sin^{-1}\Bigg(\cfrac{1}{\sqrt{x}}\Bigg)\right\}^2 & ,\quad x\geq 1 \\[2mm]
    -\cfrac{1}{4}\left(\log\cfrac{1+\sqrt{1-x}}{1-\sqrt{1-x}}-i\pi\right)^2&,
 \quad x < 1 
 \end{cases}
\qquad,\\
\notag\\
 g(x) &=
\begin{cases}
\sqrt{x^{-1}-1}\sin^{-1}\sqrt{x} &, \quad x\leq 1
\\[2mm]
    \cfrac{\sqrt{1-x^{-1}}}{2}\left(\log\cfrac{1+\sqrt{1-x^{-1}}}{1-\sqrt{1-x^{-1}}}-i\pi\right)&,
 \quad x > 1 
 \end{cases}
\qquad.
\end{align} 

\end{document}